\documentstyle[12pt,subeqnarray]{article}

\begin{document}
\newcommand{\dis}{\displaystyle}

\newcommand{\expon}{{\rm e}}
\newcommand{\D}{\Delta}
\newcommand{\be}{$$}
\newcommand{\ee}{$$}
\newcommand{\bsq}{\begin{eqnarray*}}
\newcommand{\esq}{\end{eqnarray*}}
\newcommand{\bea}{\begin{eqnarray}}
\newcommand{\eea}{\end{eqnarray}}

{\thispagestyle{empty}
\rightline{} 
\rightline{} 
\rightline{} 
\vskip 1cm

\centerline{\large \bf A Path Integral Approach to Option Pricing with Stochastic}
\centerline{\large \bf Volatility: Some Exact Results }


\vskip 2cm
\centerline{ Belal E. Baaquie \footnote{Email:phybeb@nus.edu.sg}}
\centerline{ \it Department Of Physics, National University of Singapore,}
\centerline{ \it Kent Ridge, Singapore 119260 }
\vskip 0.1in

\vskip 1cm
\centerline{\bf Abstract} \vspace{10mm}

\noindent{   The  Black-Scholes formula for pricing options on stocks and other securities has 
been  generalized  by  Merton  and  Garman  to  the  case  when  stock  volatility is 
stochastic.   The  derivation  of  the price of a security derivative with stochastic 
volatility  is  reviewed starting from the first principles of finance.  The equation  
of  Merton  and  Garman  is  then  recast  using  the  path  integration technique of 
theoretical  physics.   The  price of the stock option is shown to be the analogue of 
the  Schrodinger  wavefuction  of  quantum  mechacnics  and the exact Hamiltonian and 
Lagrangian  of the system is obtained.  The results of Hull and White are generalized 
to  the  case  when stock price and volatility have non-zero correlation.  Some exact 
results for pricing stock options for the general correlated case are derived.
}
}
\newpage

\section*{I Introduction}

 The  problem  of  pricing the European call option has been well-studied starting 
from  the  pioneering  work  of Black and Scholes [1], [2].  The results of Black and 
Scholes were generalized by Merton, Garman [3] and others for the case  of stochastic 
volatility  and  they  derived  a partial differential equation that the option price 
must satisfy.

 The  methods  of  theoretical  physics have been applied with some success to the 
problem  of  option  pricing  by Bouchaud et al [4].  Analyzing the problem of option 
pricing  from  the point of view of physics brings a whole collection of new concepts 
to  the  field  of  mathematical  finance  as  well  as  adds to it a set of powerful 
computational  techniques.   This paper is a continuation of applying the methodology 
of  physics,  in particular, that of path integral quantum mechanics, to the study of 
derivatives.

 It  should  be noted that, unlike the paper by Bouchaud and Sornette [4] in which 
the pricing of derivatives is obtained by techniques which go beyond the conventional 
approach  in  finance, the present paper is based on the usual principles of finance. 
In  particular, a continuous time random walk is assumed for the security and a risk-
free  portfolio is used to derive the derivative pricing equation; these are reviewed 
in  Sec  II.   The  main  focus  of this paper is to apply the computational tools of 
physics  to the field of finance; the more challenging task of radically changing the 
conceptual framework of finance using concepts from physics is not attempted.

 An  explicit  analytical  solution  for the pricing of the European call (or put) 
option for the case of stochastic volatility has so far remained an unsolved problem.  
Hull  and  White [5] studied this problem and obtained a series solution for the case 
when  the correlation of stock price and volatility, namely r, is equal to zero, i.e. 
$\rho=0 $ (uncorrelated).   Hull  and  White [5] also obtained an algorithm to numerically 
evaluate the option price using Monte Carlo methods for the case of $\rho \neq 0$.

 Extensive  numerical studies of option pricing for stochastic volatility based on 
the algorithm of Finucane [6] have been carried out by Mills et al [7].

 The  main  focus  and content of this paper is to study the problem of stochastic 
volatility  from  the point of view of the Feynman path integral [9]. The Feynman-Kac 
formula for the European call option is well known [2, 8].  There have also been some 
applications  of  path  integrals  in  the study of option pricing [10, 11].  In this 
paper  the  path  integral  approach  to  option  pricing  is  first discussed in its 
generality and then applied to the problem of stochastic volatility.

 The  advantage of recasting the option pricing problem as a Feynman path integral 
is  that this allows for a new point of view and which leads to new ways of obtaining 
solutions  which  are  exact, approximate as well as numerical, to the the pricing of 
options.

 In  Section II the differential equation of Merton and Garman [3] is derived from 
the  principles  of finance.  In Section III this equation is recast in the formalism 
of  quantum  mechanics.   In  Section  IV a discrete time path integral expression is 
derived  for  the  option  price  with  stochastic  volatility  which generalizes the 
Feynman-Kac  formula.   In  Section  V the path integration over the stochastic stock 
price  is  performed explicitly and a continuous time path integral is then obtained.  
And lastly in Section VI some conclusions are drawn.

\section*{ II Security Derivative with Stochastic Volatility}

 We  review  the  principles  of  finance  which  underpin  the theory of security 
derivatives,  and  in  particular that of the pricing of options.  We will derive the 
results not using the usual method used in  theoretical finance based on Ito-calculus 
(which  we  will  review  for completeness), but instead from the Langevin stochastic 
differential equation.  Hence we start from first principles.

 A  security  is  any financial instrument which is traded in the capitals market; 
this  could  be the stock of a company, the index of a stock market, government bonds 
etc.   A  security  derivative  is  a  financial  instrument which is derived from an 
underlying  security and which is also traded in the capitals market.  The three most 
widely   used  derivatives  are  options,  futures  and  forwards;  more  complicated  
derivatives can be constructed out of these more basic derivartives.

 In this paper, we analyze the option of an underlying security S. A European call 
option  on  S is a financial instrument which gives the owner of the option to buy or 
not to buy the security at some future time T > t for the strike price of K.

 At  time  $t  =  T$,
when  the option matures the value of the call option f(T) is 
clearly given by
\bsq f(T, S(T))  & =  & \left\{ 
\begin{array}{ll}   
S(T) - K, & S(T) > K \\
0, & S(T) < K \end{array} \right.   \mbox{\hspace{6cm}} (1a) \\
 &  = &  g(S)    \mbox{\hspace{10.2cm}}(1b)  \esq

The  problem  of option pricing is the following: given the price of the security 
$S(t)$ at time $t$,  what should be the price of the option $f$ at time $t < T$?  Clearly $f = 
f(t,  S(t),  K, T)$. This is a final value problem since the final value of $f$ at $t = T$ 
has been specified and its initial value at time $t$ needs to be evaluated.

 The  price  of $f$ will be determined by how the security $S(t)$ evolves in time.  In 
theoretical  finance,  it  is  common to model $S(t)$ as a random (stochastic) variable 
with  its  evolution  given  by  the stochatic Langevin equation (also called an Ito-
Wiener process) as
\be
\frac{dS(t)}{dt} = \phi S(t) + \sigma S R(t) \eqno{(2a)}
\ee
where  $R$   is the usual Gaussian white noise with zero mean; since white noise is 
assumed  to  be  independent  for  each  time  $t$,  we  have  the Dirac delta-function 
correlator given by
\be
< R_{t} R_{t^\prime}  > = \delta(t - t')   .  \eqno{(2b)}
\ee

 Note  $\phi$  is the expected return on the security $S$ and $\sigma$ is its volatility.
White noise $R(t)$ has the following important property .  If we discretize time 
$t = n\epsilon$, then the probability distribution function of white noise is given by
\be
P(R_t ) = \sqrt{\frac{\epsilon}{2 \pi}}
\expon^{-\frac{\epsilon}{2} R_t^2}    \eqno{(3a)}
\ee

 For random variable $R$  it can be shown using eqn. (3a) that
\be
R_t^2  = \frac{1}{\epsilon} + \mbox{\hspace{0.5cm}  {\rm random terms of 0(1)}}
\eqno{(3b)}   
\ee

 In  other  words,  to  leading  order  in  $\epsilon$,  the  random  variable
$R_t^2$   becomes  
deterministic.  This  property of white noise leads to a number of important results, 
and goes under the name of Ito calculus in probability theory.

 As  a  warm-up,  consider  the  case  when  $\sigma =$ constant; this is the famous case 
considered by Black and Scholes.  We have
\be
\frac{df}{dt} = \lim_{\epsilon \rightarrow 0} \frac{f(t +
\epsilon, S(t + \epsilon)) - f(t,S(t))}{\epsilon}\eqno{(4a)}
\ee

 or, using Taylors expansion

\be
\frac{df}{dt} =  \frac{\partial f}{\partial t}   + \frac{\partial f}{\partial
S} \frac{dS}{dt}  +  \frac{1}{2} \frac{\partial^2 f}{\partial S^2} \{
\frac{dS}{dt}\}^2 \epsilon  + {\rm 0}(\epsilon^{1/2}   )\eqno{(4b)}
\ee

 But
\bsq
\lbrack \frac{dS}{dt} \rbrack^2 & =&  \sigma^2 S^2 R^2  + {\rm 0}(1) \\
& = &   \frac{1}{\epsilon} \sigma^2 S^2  + {\rm 0}(1)
\mbox{\hspace{11cm}} (4c)
\esq

 Hence we have, for $\epsilon \rightarrow 0$
\be
\frac{df}{dt}    =  \frac{\partial f}{\partial t}  +  \frac{1}{2} \sigma^2
S^2 \frac{\partial^2 f}{\partial S^2}  + \phi S \frac{\partial
f}{\partial S}   + \sigma S \frac{\partial f}{\partial S}  R  \eqno{(4d)}
\ee

 Since  eqn.  (4d) is of central importance for the theory of security derivatives 
  we  also  give  a derivation of it based on Ito-calculus.  Rewrite eqn. (2a) in terms  
differentials as

\be
  dS = \phi S dt + \sigma S dz
\ee
where  the  Wiener  process $ dz = R dt$, with $R(t)$ being the Gaussian white noise.
Hence from eqn. (3b)

\be
  (dz)^2  = R_t^2  (dt)^2  = dt + {\rm   0}(dt^{3/ 2}   )
\ee
and hence 
\be
(dS)^2  = \sigma^2 S^2  dt + {\rm 0}(dt^{3/2}   ).
\ee

 From the equations for $dS$ and $(dS)^2$  given above we have
\bsq    
df & = &   \frac{\partial f}{\partial t} dt +  
\frac{\partial f}{\partial S}  dS +
\frac{1}{2} \frac{\partial^2 f}{\partial S^2}
(dS)^2  + {\rm 0}(dt^{3/2}   ) \\
& = &  (\frac{\partial f}{\partial t}   +  
\frac{1}{2} \sigma^2 S^2 \frac{\partial^2 f}{\partial S^2} ) dt +
\sigma S  \frac{\partial f}{\partial S}  dz
\esq   and using $\frac{dz}{dt}   = R$ we obtain eqn. (4d).

 The  fundamental  idea  of  Black  and  Scholes is to form a portfolio such that, 
instantaneously,  the  change  of  the portfolio is independent of the white noise $R$. 
Consider the portfolio

\be
   \pi = f -  \frac{\partial f}{\partial S}  S\eqno{ (5a) }
\ee
i.e.  $\pi$  is a portfolio in which an investor sells an option $f$ and buys  $\frac{\partial f}{\partial
S}$   amount of security $S$.  We then have from eqns. (4b) and (5a)

\bsq
\frac{d \pi}{dt} & = & \frac{df}{dt}    -  \frac{\partial f}{\partial S}
\frac{dS}{dt} \\
& = & \frac{\partial f}{\partial t}  +   
\frac{1}{2} \sigma^2 S^2 \frac{\partial^2 f}{\partial S^2}  
\esq
$$\eqno{(5b)}$$

 The  portfolio  $\pi$ has been adjusted so that its change is deterministic and hence 
is  free  from  risk  which  comes  from  the  stochastic nature of a security.  This 
technique  of  cancelling the random fluctuations of one security (in this case of $f$) 
by  another  security  (in  this  case S) is called hedging.  The rate of return on ã 
hence must equal the risk-free return given by the short-term risk-free interest rate 
$r$ since otherwise one could arbitrage.  Hence

\be
\frac{d \pi}{dt}  = r \pi \eqno{(5c)}
\ee
which yields from eqn.(5b) the famous Black-Scholes [1] equation

\be
\frac{\partial f}{\partial t}   + r S \frac{\partial f}{\partial S}   +  
\frac{1}{2} \sigma^2 S^2 \frac{\partial^2 f}{\partial S^2}= r f\eqno{(5d)} 
\ee

 Note  the  parameter  $\phi$  of eqn. (2a) has dropped out of eqn. (5d) showing that a 
risk-neutral  portfolio $\pi$ is independent of the investors expectation as reflected in 
$\phi$; or equivalently, the pricing of the security derivative is based on a risk-neutral 
process which is independent of the investors opinion.

 We  now  address the more complex case where the volatility $\sigma(t)$ is considered to 
be  a  stochastic  (random)  variable.   We  then have the following coupled Langevin 
equations (following Hull and White [5]), for $\sigma^2  = V$, 

\be
\frac{dS}{dt}  = \phi S + \sigma S R \eqno{(6a)} 
\ee
\be
\frac{dV}{dt} = \mu V + \xi V Q   \eqno{(6b)}
\ee
where  $\mu$ is the expected rate of increase of the variance $V$, $\xi$ its volatility and 
$ R$ \& $Q$ are correlated Gaussian white noise with zero means and with correlators
\be
< R_t R_{t^\prime}   > = < Q_t  Q_{t^\prime}   > = \delta(t - t') \eqno{(6c) }
\ee
and
\be
< R_t Q_{t^\prime}   > = \rho \delta(t - t')  \eqno{(6d) } 
\ee
with  $\rho^2  < 1$ being the reflection of the  correlation between $S$ and $V$.

 Note  that  one  can  choose  a  different  process  from eqn. (6) for stochastic 
volatility  as  has  been  done  by  various  authors  [5] but the main result of the 
derivation is not affected.

 The  procedure  for  deriving  the  equation  for  the price of an option is more 
involved  when volatility is stochastic.  Since volatility $V$ is
{\it not} a traded security 
this  means that a portfolio can only have two instruments, namely $f$ and $S$.  To hedge 
away  the  all  fluctuations  due  to  both S and V, we need at least
{\it three} financial 
instruments  to  make  a  portfolio  free  from the investors subjective preferences.  
Since  we  have  only two instruments from which to form our portfolio we will not be 
able to completely avoid the subjective expectations of the investors.

 Consider  the  change  in the pricing of the option in the presence of stochastic 
volatility; using eqns (6a) and (6b) we have from Taylors expansion

\bsq
\frac{df}{dt} &= & \lim_{\epsilon \rightarrow 0} \frac{1}{\epsilon} 
\{ f(t + \epsilon, V(t + \epsilon), S(t+\epsilon) - f(t, V(t), S(t))  ) \}   \\ 
&   = & \frac{\partial f}{\partial t} + \frac{\partial f}{\partial S}
(\phi S + \sigma S R)    +  \frac{\partial f}{\partial V}  (\mu V + \xi
V Q)  \\
& & + \frac{\epsilon }{2} \left\{\frac{\partial^2 f}{\partial S^2}
[\frac{dS}{dt}]^2 + \frac{\partial^2 f}{\partial V^2} [\frac{dV}{dt}]^2
 + 2 \frac{\partial^2 f}{\partial S \partial V} \frac{dV}{dt}
\frac{dS}{dt} \right\} +
{\rm 0}(\epsilon^{1/2}   )   \mbox{\hspace{4cm}}  (7a)
\esq

Using
$$
R_t^2   =   Q_t^2  = + {\rm 0}(1), $$
$$
R_t Q_t  =   \frac{\rho}{\epsilon} + {\rm 0}(1)  
\eqno{(7b)} $$
we have, similar to eqn. (4a), for $\epsilon \rightarrow 0$
$$   
\frac{df}{dt}  =  \frac{\partial f}{\partial t} + \phi S \frac{\partial f}{\partial S}
 +  \mu V \frac{\partial f}{\partial V}  + \frac{1}{2} \left[
\sigma^2 S^2 \frac{\partial^2 f}{\partial S^2} + \zeta^2 V^2 \frac{\partial^2 f}{\partial
V^2} + \xi \sigma^3 S \frac{\partial^2 f}{\partial V \partial S}
\right] $$
$$  \mbox{\hspace{3cm}} + \sigma S \frac{\partial f}{\partial S} R + \epsilon V \frac{\partial f}{\partial
V} Q  \eqno{(7c)}
$$

 Define the deterministic and stochastic terms by
\be
\frac{df}{dt} \equiv f\omega + f \alpha_1 R + f \alpha_2  Q    \eqno{(7d) }
\ee

 Consider  a portfolio which consists of $\theta_1(t)$
amount of option $f$ and $\theta_2 (t)$ amount of stock $S$, that is
\be \pi(t) = \theta_1(t)f(t) + \theta_2(t)S(t)  \eqno{(8a) }
\ee

 Then
\be
\frac{d\pi}{dt}  =  \theta_1 \frac{df}{dt} + \theta_2 \frac{dS}{dt}
+  \frac{d\theta_1}{dt}   f +   \frac{d\theta_2}{dt}  S    \eqno{ (8b)} 
\ee

We consider a self-replicating portfolio in which no cash flows in or out and the 
quantities $\theta_1 (t)$ and $\theta_2 (t)$ are adjusted accordingly.  It then follows that 
\be
\frac{d\pi}{dt} = \theta_1 \frac{df}{dt}    + \theta_2 \frac{dS}{dt} \eqno{(8c)}
\ee

 Hence
\be
\frac{d\pi}{dt} = \theta_1 f \omega + \phi \theta_2  S +(\alpha_1 R + \alpha_2 Q) \theta_1 f
+ \theta_2 S \sigma R    \eqno{(8d)} 
\ee

 We choose the portfolio such that, in matrix notation
\be
\left[ \begin{array}{ll}  
\alpha_1 & \sigma \\
\alpha_2 & 0 
\end{array} \right] 
\left[ \begin{array}{l}  \theta_1 f \\
\theta_2 S \end{array} \right]  = 0  \eqno{(8e)} \ee
which yields
\be  \frac{d\pi}{dt}  = \theta_1 f \omega + \phi \theta_2 S  \eqno{(9a)} 
\ee

 Note that all the random terms from the rhs of eqn. (8d) have been removed by our 
choice  of  portfolio which satisfies the constraint given in eqn. (8e).  Since  
$\frac{d\pi}{dt}$  is 
now deterministic, the absence of arbitrage requires 
\be
\frac{d\pi}{dt} = r \pi  \eqno{ (9b)} 
\ee
or, from eqns. (8d) and (9a) 
\be
(\omega - r) \theta_1 f + (\phi - r) \theta_2 S = 0  \eqno{(9c) }
\ee
We  hence  have from eqns. (8e) and (9c) that eqn. (9c) 
above can be satisfied in 
general by making the following choice $\omega$ and $\phi$, namely
\be
\lbrack \omega - r , \phi - r \rbrack = [\lambda_1 (t) , \lambda_2
(t)]   \left[ 
\begin{array}{ll}  
\alpha_1 & \sigma \\
\alpha_2 & 0 
\end{array} \right] 
\ee  
where $\lambda_1$  and $\lambda_2$  are arbitrary.  In components, we have from above

$$
 \omega - r  =   \lambda_1 \alpha_1  + \lambda_2 \alpha_2    
\eqno{(9d)} $$
$$ 
\phi - r = \lambda_1 \sigma  \eqno{(9e)}
$$

 Eliminating  $\lambda_1$  yields using eqn. (9e) yields from eqn. (9d)
\be
f \omega - r f = \left[ \frac{\phi - r}{\sigma} \right]  + \lambda_2 f \alpha_2 
\ee
and  hence  from eqns. (7c) and (7d) we obtain the Merton and Garman [3] equation 
given by
\be
\frac{\partial f}{\partial t} + \frac{1}{2} \left[ \sigma^2 S^2
\frac{\partial^2 f}{\partial S^2} + 2 \rho \xi \sigma^3 S
\frac{\partial^2 f}{\partial S \partial V} + \xi^2 V^2 \frac{\partial^2
f}{\partial V^2} \right] - rf  = - rS \frac{\partial f}{\partial
S} + \left[ \lambda_2 - \mu \right] V \frac{\partial f}{\partial V}
\eqno{(10a)}
\ee

 Note  the  fact that both $\phi$ and $\lambda_1$  have been eliminated from (10a) since we could 
hedge  for  the fluctuations of $S$ by including it in the portfolio $\pi$.  In order words 
the  subjective view of the investor visa vis security $S$ has been removed.   However, 
the  appearance  of  $\lambda_2$   in  eqn.  (10a) reflects the view of the investor of what he 
expects for the volatility of $S$.  Since $V$ is not traded, the investor can't execute a 
perfect  hedge  against  it  as  was  the  case for constant volatility and hence the 
pricing of the option given in eqn. (10a) has a subjective factor in it.

 It  has  been argued by Hull and White [5] that for a large class of problems, we 
can consider l  to be a constant which in effect simply redefines $\mu$ to be $\mu -
\lambda_2$  .

 Hence we have

\be  \frac{\partial f}{\partial t} + \frac{1}{2} \left[ \sigma^2 S^2
\frac{\partial^2 f}{\partial S^2} + 2 \rho \xi \sigma^3 S
\frac{\partial^2 f}{\partial S \partial V} + \xi^2 V^2 \frac{\partial^2
f}{\partial V^2} \right] - rf  = - rS \frac{\partial f}{\partial
S} - \mu \sigma^2 \frac{\partial f}{\partial V}
\eqno{(10b)}
\ee

 Eqn.  (10)  above  is  an  equation  for  {\it any} security derivative with stochastic 
volatility.   Whether  it  is  an  European  option or an Asian option or an American 
option  is determined by the {\it boundary conditions} that are imposed on $f$; in particular 
the boundary condition given in eqn. (1) is that of an European call option.

\section*{III Black-Scholes-Schrodinger Equation for Option Price}

 We  recast  the  Merton   and  Garman equation given in eqn. (10) in terms of the 
formalism of quantum mechanics; in particular we derive the Hamiltonian for eqn. (10) 
which is the generator of infinitesimal translations in time.  Since both $S$ and $V$ are 
random  variables which can never take negative values, we define new variables $x$ and 
$y$ 

\bsq   
S & = & \expon^x   ,  - \infty < x <  \infty   \mbox{\hspace{11cm}} (11a)  \\
V & = & \expon^y   ,  - \infty <  y <  \infty .  \mbox{\hspace{11cm}} (11b)  
\esq

 From  eqn.  (10b),  we then obtain the Black-Scholes-Schrodinger equation for the 
 price of security derivative as

\be 
\frac{\partial f}{\partial t}  = (H+r)f \eqno{(12a)}
\ee 
where $f$ is the Schrodinger wave function.  The ``Hamiltonian' $H$ is a differential 
operator given by
\be
H   =   -\frac{\expon^Y}{2} \frac{\partial^2}{\partial x^2}
+ ( \frac{1}{2} \expon^Y  - r) \frac{\partial}{\partial x}
- \xi \rho \expon^{Y/2}   \frac{\partial^2}{\partial x \partial y} -
\frac{\xi^2}{2} \frac{\partial^2}{\partial y^2} +
(\frac{1}{2} \xi^2 - \mu) \frac{\partial}{\partial y}
\eqno{(12b) }
\ee

 From eq. (12a) we have the formal solution
\be
f = e^{t(H + r)} \tilde{ f}    \eqno{(13)}  
\ee
where $\tilde{ f}$ is fixed by the boundary condition.  Note from eqn. (13) that option price $f$ 
seems  to be unstable being represented by a growing exponential; however, as will be 
seen later since the boundary condition for $f$ is given at final time $T$ eqn. (13) will 
be converted to a decaying exponential in terms of remaining time $ç = T - t$.

 Let  $p(x, y, \tau |x^\prime, y^\prime)$, be the conditional probability that, given security price 
$x^\prime$  and  volatility  $y^\prime$ at time $\tau = 0$, it will have a  value of $x$ and volatility $y$ at  
time  $t$.   We  have  the  boundary condition at $\tau = 0$ given by Dirac delta-functions, 
namely
\be
 p(x, y, 0 | x^\prime, y^\prime) = \delta(x^\prime  - x)\delta(y^\prime  - y)  \eqno{ (14)}
\ee

The  derivative  price  is  then given, for $0 \leq t \leq T$, 
by the Feynman-Kac formula 
Refs [5, 8] as
\be
f(t; x, y) = \expon^{-(T - t)r} \int_{-\infty}^{+\infty} 
dx^\prime p(x, y, T-t³x^\prime)g(x^\prime)   \eqno{(15a)}
\ee
where
\be
p(x, y,  T-t³x^\prime) =\int_{-\infty}^{+\infty}  dy^\prime p(x, y,
T-t³x^\prime, y^\prime) \eqno{(15b)}
\ee
 Note  that  integration over $y^\prime$ 
is decoupled from the  function $g(x^\prime)$ due to eqn. 
(1).  It follows from eqn. (14) that $f(t; x, y)$ given in (15a) 
satisfies the boundary condition given in eqn. (1), i.e.

\be   f(T, x, y) = g(x)   \eqno{(15b)   }
\ee
We rewrite eqns. (15a) and (15b) in the  notation of quantum
mechanics.  The function  $ f(x)$  can be thought of as an infinite 
dimensional vector $|f>$ of a function space with components 
$ f(x)  =  <  x|f  >$  and  $f^\ast(x)  =  <  f|x  >$  
(where * stands for complex conjugation).  
For $- \infty \leq x \leq \infty$  
the bra vector $< x >$ is the basis of the function space 
and the ket vector $|x >$ is its dual with normalization
\be < x|x' > = \delta(x - x').   \eqno{(16) }
\ee

The scalar product of two functions is given by
\bsq
< f|g > & \equiv &dx f^\ast(x) g(x) \\
&   =  & < f|\{\int_{-\infty}^{\infty} \int dx |x><x| \} |g>
\esq
and yields the completeness equation
\be
I = \int_{-\infty}^{\infty}    dx |x >< x| \eqno{(17a)}
\ee
where $I$ is the identity operator on function space.  
For the case of stochastic stock 
price and stochastic volatility, we have 
$$   I =  \int_{-\infty}^{\infty}   dx dy |x, y >< x, y| \eqno{(17b)}
$$ 
where $|x , y > \equiv |x > \otimes |y >$.

 From eqn (13) we have, in Dirac's notation 
$$   |f, t > = \expon^{t(H + r)} |f, 0 >  $$
and boundary condition given in eqn. (1b) yields 
$$   |f, T > = \expon^{t(H + r)} |f, 0> = |g>. $$

Hence
$$  
|f, t > = \expon^{(T - t)(H + r)} |g> \eqno{(18a)} 
$$
or, more explicitly
\bsq  f(t, x, y) & = & < x, y | f, t > \\
&  = &  \expon^{-r(T - t)} <x, y| \expon^{-(T - t)H}  |g>
\mbox{\hspace{8cm}} (18b)
\esq 
or, using completeness equation (17b)
$$ 
f(t,x,y) = \expon^{-r(T - t)}  \int_{-\infty}^{\infty}   dx\prime
dy^\prime <x, y|  \expon^{-(T - t)H}   |x^\prime, y^\prime> g(x^\prime)
\eqno{(19a) }
$$
which yields for remaining time $\tau = T - t$, from eqn. (15b)
$$
p(x, y, \tau | x^\prime, y^\prime) = <x, y|\expon^{- \tau H}  |x^\prime, y^\prime>.
\eqno{(19b)}  
$$
Note  remaining  time  runs backwards, i.e. when $\tau = 0$, 
we have $\tau = T$ and when $\tau = T$, 
real  time  $t  =  0$; as mentioned earlier, expressed in terms 
of remaining time $\tau$ the 
derivative price is given by a decaying exponential as given in eqn. (19b).

Before  tackling  the  more  complicated  case  of  
option  pricing  with  stochastic  
volatility  I  apply  the  formalism developed so far 
to the simpler case of constant 
volatility;  the  detailed derivation is given in Appendix 1.  
The well known results 
of Black and Scholes are seen to emerge quite naturally in this formalism.

\section*{IV. The Discrete-Time Feynman Path Integral}

 The  Feynman  path  integral  [9]  
is a formulation of quantum mechanics which is 
based on functional integration.  In particular, 
the Feynman path integral provides a 
functional integral realization of the conditional probability 
$p(x, y, \tau|x^\prime, y^\prime)$.  To 
obtain  the  path  integral  we discretize time $\tau$ 
into $N$ points with spacing $ \epsilon= \tau/N$.  
Then, from (19b), 
$$ p(x, y, \tau|x^\prime, y^\prime) = 
\lim_{N \rightarrow \infty} <x, y|\expon^{-\epsilon H} \cdots
\expon^{-\epsilon H}    |x^\prime, y^\prime>  . \eqno{(20)}   
$$
Hence forth, we will always assume the $N \rightarrow
\infty$ limit.

 Inserting  the completeness equation 
for $x$ and $y$, namely eqn (17b), $(N - 1)$ times 
in eqn.(20) yields
$$ 
p(x, y, \tau|x^\prime, y^\prime) = 
( \prod_{i = 1}^{N-1} \int  dx_i dy_i )  
\prod_{i = 1}^{N}  <x_i ,y_i |\expon^{- \epsilon H}  
| x_{i -1},y_{i -1} >    \eqno{ (21a)}  
$$
with boundary conditions
$$ x_0  = x^\prime, \mbox{\hspace{1cm}}    y_0  = y^\prime \eqno{(21b)}  
$$
$$
x_N  = x,  \mbox{\hspace{1cm}}    y_0  = y.   \eqno{(21c)}  
$$

 We  show  in the Appendix 2 that for the H given by eqn
(12b) we have the Feynman  relation given in (A9) and (A13)
$$
< x_i , y_i |\expon^{-\epsilon H}  |x_{i -1}, y_{i -1} > 
= \frac{1}{2 \pi \epsilon} 
\frac{\expon^{- y_i/2}}{\xi\sqrt{1 - \rho^2}} \expon^{\epsilon L(i)} 
\eqno{(22a)}  
$$
where  the  `lagrangian'  L is given in eqn (A15), 
for $\delta x  \equiv x_i  - x_{i -1}$    and 
$\delta y  \equiv  y_i  -  y_{i - 1} $   as
$$
L(i) =  -\frac{1}{2 \xi^2} ( \frac{\delta y_i}{\epsilon} + \mu
-\frac{1}{2} \xi^2 )^2 
$$
$$
- \frac{\expon^{-y_i}}{2(1-\rho^2)} \{ \frac{\delta x_i}{\epsilon}
+ r - \frac{1}{2} \expon^{y_i} -  \frac{\rho}{\xi} \expon^{y_i/2}
(\frac{\delta y_i}{\epsilon}    + \mu -  \frac{1}{2} \xi^2) \}^2  + {\rm
O}(\epsilon)   \eqno{(22b)}
$$

For $\epsilon \rightarrow 0$, we have, as expected 
$$
< x_i , y_i |\expon^{-\epsilon H}   |x_{i -1}   , y_{i-1} > 
= \delta(x_i  - x_{i-1}   )\delta(y_i  - y_{i -1}) + {\rm O}(e)
\eqno{(23)}
$$
and the prefactors to the exponential on the rhs of (22a) 
ensure the correct limit.

 We form the `action' $S$ by 
$$ S = \epsilon  \sum_{i =1}^{N} L(i) + {\rm O}(e).
\eqno{(24)}  
$$
Note $S$ is quadratic in $x_i$  and non-linear in the $y_i$  
variables.

Hence, from eqns. (20), (21), (22) and (24) we have the path integral
$$
p(x, y, \tau | x^\prime) =\int_{-\infty}^{\infty}  
dy^\prime p(x, y, \tau | x^\prime, y^\prime)
\eqno{(25a)}
$$
$$
\mbox{\hspace{2cm}} = \lim_{N \rightarrow \infty} \int   
DXDY \expon^S \eqno{(25b)} 
$$
where, for $ \epsilon = \tau/N$
$$   \int  DX =
\frac{\expon^{-Y_N/2}}{\sqrt{2 \pi \epsilon (1 - \rho^2)}} 
\prod_{i = 1}^{N -1} \int_{-\infty}^{\infty}  \frac{dx_i
\expon^{-y_i/2}} {\sqrt{2 \pi \epsilon (1 - \rho^2)}} 
\eqno{(25c) }
$$
and  
$$  
\int DY = \prod_{i = 1}^{N -1} \int_{-\infty}^{\infty}
\frac{dy_i}{\sqrt{2 \pi \epsilon \xi^2}}.
\eqno{(25d)}
$$

Note we have included an extra $dy_0  = dy^\prime$ 
integration in $DY$ given in (25d) due to 
the integration over $y^\prime$ in eqn (25a).

Eqn  (25)  is  the  path  integral  for  stochastic  stock  price 
with stochastic 
volatility  in  its full generality.  To evaluate expressions 
such as the correlation 
of $S(t)$ with $V(t^\prime)$, the path integral in eqn (25) has to be used.

Since  the  action $ S$  is  quadratic  in  the $ x_i$,  
the path integral over the $x_i$  
variables in eqn. (25) can be done exactly.  
The details are given in Appendix 3.

  From eqn (A21) we obtain 
$$  p(x, y, \tau | x^\prime) = \int DY 
\frac{\expon^{S_0 + S_1}}{\sqrt{2 \pi \epsilon (1 - \rho^2) \sum_{i =
1}^{N} \expon^{y_i}}}  \eqno{(26a)  }
$$
where the first term in eqn. (22b) gives
$$ S_0  = -\frac{\epsilon}{2 \xi^2} \sum_{i = 1}^{N} (\frac{\delta
y_i}{\epsilon} + \mu  - \frac{\xi^2}{2}  )^2
\eqno{(26b)}  
$$
and $S_1$  is the result of the $DX$ 
path integration given by eqn (A21b) as 
$$
S_1  = -\frac{1}{2(1 - \rho^2) \epsilon \sum_{i = 1}^{N} \expon^{y_i}}
\{ x - x^\prime  + \epsilon \sum_{i = 1}^N (r -  \frac{1}{2} \expon^{y_i}  )
$$
$$\mbox{\hspace{6cm}}    -\frac{\rho}{\xi}  \sum_{i =1}^{N}
\expon^{y_i/2}  [\delta y_i  + \epsilon (\mu - \frac{\xi^2}{2})] \}^2.
\eqno{ (26c)  }
$$

Extensive  numerical  studies  of  the  pricing  of 
European call option has been 
carried out in [13] based on eqns. (26a-c).

\section*{V.  Continuous Time Feynman Path Integral}

We first derive the continuum limit of the 
Black-Scholes constant volatility case 
before  analyzing  the  more  complex  case of stochatic volatility.  
The $DX_{BS}$   - path 
integral for the Black-Scholes case , from eqn (A5b), has a measure 
$DX_{BS}   = [\frac{1}{2 \pi \epsilon \sigma^2}]^{N/2}
\prod_{i = 1}^{N} dx_i$   
which is essentially the measure for the flat space ${\bf R^N}$; 
we can hence take the 
$N \rightarrow \infty$ for $DX$   
and obtain a well defined continuous-time path integral [9].  From eqn 
(A5a) we obtain taking the continuum limit of $\epsilon \rightarrow  0 $
$$   S_{BS}   =\int_{0}^{\tau}    dt L_{BS}   
= -\frac{1}{2 \sigma^2} \int_0^{\tau}   dt (\frac{dx}{dt} + r -
\frac{1}{2} \sigma^2 )^2  
$$
with boundary conditions $x(0) = x^\prime$ 
and $x(\tau) = x$.  We see from above that the Black-Scholes case of 
constant volatility corresponds to the evolution of a free quantum 
mechanical particle with mass given by $1/\sigma^2$ .  
This yields the Black-Scholes result
$$
p_{BS}  (x, \tau | x^\prime) = < x | \expon^{- \tau H_{BS}}
|x^\prime > = \int  
DX_{BS}   \expon^{S_{BS}}   
$$
The exact result for $p_{BS} (x, \tau | x^\prime)$ is given in
eqn (A4).

Eqn. (26) provides a mathematically rigorous basis for taking the 
$N \rightarrow \infty$ limit for 
the  case of stochastic volatility.  
On exactly performing the  $\int DX$-path integral, the 
remaining  $\int DY$ - path integral has a measure $DY = [\frac{1}{2\pi
\epsilon \xi^2} ]^{N/2} \prod_{i = 1}^{N} dy_i $  which, just as in 
the  Black-Scholes case, is essentially the measure for the 
flat space ${\bf \rm R^N}$  (unlike the 
nonlinear  expression in eqn. (25c)); we can hence take the 
$N \rightarrow \infty$ for $DY$ and for 
$S_0  + S_1$  and obtain a well defined continuous-time path integral.

We take the limit of $\epsilon \rightarrow  0$; 
we have, $t = i \epsilon $,  $\epsilon \sum_{i =1}^{N} \expon^{Y_i} 
\rightarrow   \int_{0}^{\tau} dt \expon^{Y(t)}  \equiv \tau  w $, 
and 
$\frac{1}{\epsilon} \delta y_i \rightarrow  \frac{dy}{dt}$.  
Hence, from (26) we have
$$
S = S_0  + S_1 \eqno{(27a)}   
$$
$$
S_0  = -\frac{1}{2 \xi^2}  \int_0^\tau
 dt ( \frac{dy}{dt}  + \mu -  \frac{1}{2} \xi^2)^2
\eqno{(27b)}
$$
$$   
S_1    =   -\frac{1}{2(1 - \rho^2)  w } \{x - x^\prime + r \tau 
-  \frac{1}{2} \int_{0}^{\tau} dt \expon^{Y(t)}    
$$
$$
\mbox{\hspace{3cm}}   + 
\frac{2\rho}{\xi} (\expon^{Y(0)/2} - \expon^{Y/2}) 
- \frac{\rho}{\xi}(\mu - \frac{\xi^2}{2} ) \int_0^\tau  dt
\expon^{Y(t)/2} \}^2 \eqno{(27c)}
$$
with boundary value
$$y(\tau) = y   \eqno{ (27d)   }
$$
and
$$  p(x, y, \tau | x^\prime) =  \int DY \frac{\expon^S}{\sqrt{2\pi(1 -
\rho^2)\tau  w }}.  \eqno{(28)}
$$

Note  taking  the  continuum  limit is possible only after the discrete
$\int DX$ path 
integration  has  been  performed since the nonlinear measure 
given by (25c) does not 
have  a  finite  continuum  limit.   All  correlation  functions  
$< S(t)V(t^\prime)>$ can be 
obtained from the continuum limit of the discrete correlators 
$< \expon^{X_n}   \expon^{Y_m}  >$.

For  the  case of $\rho
 = 0$, 
to recover the Black-Scholes formula, set $w = \sigma ^2$, 
where 
$\sigma=\expon^{Y/2}$   
is the volatility at time $t = 0$.  
As has been noted by Hull and White [5], for 
the  case  of $\rho= 0$ the conditional probability depends on 
stochastic volatility $y(t)$ 
only through the combination $ w  = \frac{1}{\tau} \int_0^\tau 
\expon^{Y(t)/2}  dt$.  For $\rho \neq 0$, we see from eqn. (27c) 
that $ p(x, y, \tau³x^\prime)$ 
now depends on $ w $ as well as on $\dis u = \frac{1}{\tau}
\int_0^\tau   \expon^{Y(t)/2}  dt$  and $\expon^{Y(0)/2}$.
We hence have from (27) and (28) 
$$ 
p(x, y, \tau | x^\prime) = \int_0^\infty   dw \;\; du\;\; dv\;\;  
\frac{\expon^{S_1(u,v,w)}}{\sqrt{2 \pi(1 - \rho^2) \tau w}} g(u,v,w)
\eqno{(29a)   }
$$
where
$$  S_1 (u,v,w) = - \frac{1}{2(1 - \rho^2) \tau w} \{   
x - x^\prime + r\tau - \frac{\tau}{2} w +  
\frac{2 \rho}{\xi} (v - \expon^{Y/2}   )
$$
$$ \mbox{\hspace{3cm}}    -  \frac{\rho}{\xi} 
( \mu - \frac{\xi^2}{2} )u \}^2  .  \eqno{(29b)}   
$$
and $g(u,v,w)$ is given by the path integral
$$
g(u,v,w) = \int  DY \expon^{S_0}   \delta \{  
w - \frac{1}{\tau} \int_0^\tau \expon^{Y(t)} dt \}  
\delta \{v - \expon^{Y(\tau)/2} \}  
$$
$$ \mbox{\hspace{5cm}} \delta \{ u - \frac{1}{\tau} \int_0^\tau 
\expon^{Y(t)/2} dt \} \eqno{(29c)}   
$$

From  eqn (29c) we see that $g(u,v,w)$ 
is the probability density for $u, v,$ and $w$.  
The  path integral for $g(u,v,w)$ 
is nonlinear and cannot be performed exactly.  We see 
that  $p(x,  y,  \tau|x^\prime)$ 
is the weighted average of the integrand 
$\expon^{S_1(u,v,w)}/\sqrt{2\pi(1-\rho^2 )\tau w}$  
with respect to $g(u,v,w)$.

Note  $g(u,v,w)$  has  the  remarkable  property  that  
it  is  independent  of $\rho$.  
Following  Hull  and  White  [5] one can expand the 
integrand in (29a) in an infinite 
power series in $u, v, w$ and reduce the evaluation of 
$p(x, y, \tau | x^\prime)$ to finding all the 
moments of $u, v$ and $w$; in other words we need to evaluate 
$$< u^n w^m v^p  > \equiv \int_0^\infty  
du\;\; dw\;\; u^n\;\;  w^m\;\;  v^p\;\;  g(u,v,w)  \eqno{ (30a)}
$$
$$ 
\mbox{\hspace{3cm}} = \int  DY [ \frac{1}{\tau} \int_0^\tau \expon^{Y(t)/2} dt]^n  
[ \frac{1}{\tau} \int_0^\tau \expon^{Y(t)}  dt]^m  \expon^{pY(0)/2} 
\expon^{S_0}    \eqno{(30b)}
$$

The path integral for $< u^n w^m v^p  >$ can be performed exactly.  Rewrite (30b) as
$$
< u^n w^m v^p  > = \frac{1}{\tau^{n + m}} \int_{0}^{\tau}
dt_1 \cdots dt_n  dt_{n+1} \cdots dt_{n + m}    Z(j,y,p) \eqno{(30c)}   
$$
where
$$  
Z(j,y,p) = \int  DY \exp[\int_0^\tau  dt j(t)y(t)] \expon^{pY(0)/2}
\expon^{S_0}   \eqno{(31a)}
$$
and from eqns. (30b) and (30c) we have
$$   j(t) = \frac{1}{2} \sum_{i =1}^{n} \delta (t - t_i ) 
+ \sum_{i = n+1}^{n + m} \delta (t - t_i ) +   \frac{p}{2} \delta(t)
\eqno{(31b)}
$$

$$ \mbox{\hspace{3cm}} \equiv   
\sum_{i =1}^{n + m} a_i \delta(t - t_i ) + \frac{p}{2}  \delta(t).   
\eqno{(31c)}   
$$
 The path integral for $Z(j,y,p)$ 
is evaluated exactly in Appendix 3 and yields 
$$
< u^n w^m v^p  > = \sigma^p \frac{\expon^{\sum_i a_i Y}}{\tau^{n + m}} 
\prod_{i = 1}^{n + m} \int_0^\tau dt_i  \expon^F
\eqno{(32a)}
$$
where from eqns (A41a), (A41b) and (31c) we have, after some simplifications
$$
F = (\mu -  \frac{1}{2} \xi^2 ) \sum_{i =1}^{n + m} a_i t_i  
+ \xi^2 \sum_{i,j = 1}^{n + m}  a_i a_j \theta(t_i  - t_j )t_j  +
F^\prime    \eqno{(32b)}
$$
with
$$
F^\prime =  \frac{1}{2} p \tau (\mu -  \frac{1}{2} \xi^2 ) 
+  \frac{1}{2} p \xi^2 \sum_{i = 1}^{n +m} a_i t_i  + \frac{1}{8} 
p^2 \xi^2 \tau.  \eqno{(32c)}
$$

Note that in obtaining eqn (32) we have used the identity that 
$\theta(0) = \frac{1}{2}$, as given 
in eqn. (A37a) and reflects the identity 
$\int_0^\tau  dt \delta (t - \tau) = \frac{1}{2} $.  
All the $t_i$  integrations 
in  (32)  can  be  performed  exactly  since the exponent is 
linear in the $t_i$ 's.  The 
expression  (32) for $<u^n w^m v^p >$ 
generalizes the result of Hull and White [5] as we have 
an  exact  expression  for  all the moments of $u, v$ and $w$ 
as well as for their cross-correlators.

We explicitly evaluate the first few moments using eqn (32).
$$
< w > =  \frac{\expon^Y}{\tau} \int^\tau_0 dt \expon^{\mu t} 
$$
$$ = \frac{V}{\mu \tau} (\expon^{\mu \tau} -1) \eqno{(33)}    
$$
where $V = \expon^Y$;
$$
< w^2  > = \frac{2 \expon^{2Y}}{\tau^2}  \int_0^\tau dt_1  \int_0^{t_1}
   dt_2  \expon^{\mu(t_1 + t_2)} \expon^{\xi^2 t_2}    
$$

$$ \mbox{\hspace{3cm}}   = \frac{2V^2}{\tau^2} \left[ 
\frac{\expon^{(2 \mu + \xi^2)\tau}}{(\xi^2 + \mu)(2 \mu + \xi^2)} - 
\frac{\expon^{\mu \tau}}{\mu(\mu + \xi^2)} + \frac{1}{\mu(2 \mu +
\xi^2)} 
 \right].   \eqno{(34)}
$$

We evaluate $<w^3>$ for the case of $\mu = 0$; we have from eqn (32)
$$
<w^3> =  \frac{6 \expon^{3Y}}{\tau^3} \int_0^\tau dt_1 \int_0^{t_1} dt_2
\int_0^{t_2} dt_3 \expon^{\xi^2 (t_2 + 2t_3)} 
$$

$$ \mbox{\hspace{2cm}} = \frac{V^3}{3 \xi^6 \tau^2} (\expon^{3 \xi^2
\tau} - 9 \expon^{\xi^2 \tau} + 8 + 6 \xi^2 \tau )  \eqno{ (35)}
$$

Eqns  (33),  (34)  and  (35)  agree  exactly  with the results 
stated (without derivation)  in  Hull  \&  White [5].  It is reassuring 
to see that two very different 
formalisms  agree  exactly  and  this  increases ones confidence 
in the path integral 
approach.

We  compute  a  few  moments that have not been given in Hull and 
White [5].  We further have (recall $\sigma = \expon^{Y/2}$   ) 

$$
< u > = \frac{\expon^{Y/2}}{\tau} \int_0^\tau dt \expon^{(\mu/2 -
\xi^2/8)t}
$$

$$ \mbox{\hspace{3cm}}
= \frac{2 \sigma}{\tau (\mu - \xi^2/4)} \left[
\expon^{(\mu/2 - \xi^2/8)\tau} -1 \right].   \eqno{(36a)}
$$

$$
<u^2> = \frac{2 \expon^Y}{\tau^2} \int_0^\tau dt_1 \int_0^{t_1} dt_2 
\expon^{\mu (t_1 + t_2)/2} \expon^{-\xi^2 (t_1 - t_2)/8}  \eqno{(36b)}
$$

$$
= \frac{4V}{\tau^2} \left[ 
\frac{\expon^{(\mu/2 - \xi^2/8)\tau}}{(\mu^2 + \xi^2/4)(\mu - \xi^2/8)}
- \frac{2 \expon^{(\mu/2 - \xi^2/4) \tau}}{(\mu + \xi^2/4)(\mu -
\xi^2/2)} - \frac{1}{(\mu - \xi^2/2)(\mu - \xi^2/8)}
\right] \eqno{ (37) }
$$

and lastly, for $a_1  = \frac{1}{2}$, $a_2  = 1$ 
in  eqn (31c), we have
$$
< uw > =   \frac{2 \expon^{3Y/2}}{\tau^2} \int_0^\tau  dt_1  
\int_0^{t_1}  dt_2  \expon^{\mu(t_1/2 + t_2)} \expon^{\xi^2 (t_1 - 6
t_2)/8} 
$$

$$ 
= \frac{4 \sigma V}{\tau^2} \left[ 
\frac{\expon^{(3\mu/2 + 3\xi^2/8)\tau}}{3(\mu + \xi^2/2)(\mu +\xi^2/4)}
- \frac{\expon^{(\mu/2 - \xi^2/8) \tau}}{(\mu + \xi^2/2)(\mu^2 -
\xi^2/4)} + \frac{1}{3(\mu - \xi^4/16)}
\right].\eqno{ (38)  }
$$

 All  the  other  moments  $<  u^n w^m v^p   >$ 
can similarly be evaluated, which in turn 
yields an infinite series solution for $p(x, y, \tau | x^\prime)$ 
in eqn (29a).

\section*{VI. Conclusion}

The  complete  information regarding the dynamics of how stock price
$S(t)$ and its 
volatility  $V(t)$  evolve,  their  cross-correlators  
as well as their fluctuations is 
given  by  the  discrete  time  path  integral given in eqn. (25).  
In particular, to 
numerically  study  correlators  such  as  $<  \expon^{X_n}
\expon^{Y_m}  >$  we need to start from the path 
integral given in eqn. (25).

However,  if  one  is interested solely in the price of the option, 
one needs to 
only  determine $p(x, y, \tau | x^\prime, y^\prime)$ 
and in this case the simplified discrete-time path 
integral  obtained  in  eqn.  (26) should be used as 
the starting point for numerical 
studies.  A detailed numerical simulation based on the 
results of this paper has been 
carried  out  in  ref  [13]; in particular, it has been shown 
that implied volatility 
smile  and  frown can be obtained for different values 
of the correlation parameter $\rho$
as well as for varying values of $S(t)$.

In  summary,  we  reformulated  the  option  pricing  problem 
in the language of 
quantum  mechanics.   We then obtained a number of exact results 
for the option price 
of  a  security  derivative  using  the  path integral method.  
Stochastic volatility 
introduces a high order of nonlinearity in the option pricing 
problem and needs to be 
studied using approximate and numerical techniques.

The  continuous  time  path  integral  given  in  eqn. (29c) 
is a nonlinear path 
integral  which  has  no  known  exact solution.  Various 
techniques could be used to 
obtain new approximations for g(u,v,w) relevant for special applications.

Numerical  algorithms  based  on  the  path integral 
provide a wide range of new 
numerical algorithms which go beyond the usual binomial tree 
and its generalizations.  

The comparative efficiency of the different algorithms is being
studied.  The Feynman 
path  integral  has  been  numerically  studied using many algorithms 
and which could 
prove useful in addressing problems in finance.

\centerline{\bf Acknowledgements}

I would like to thank Dr Lawrence Ma, Dr Toh Choon Peng and Dr Ariff 
Mohamed for 
having  introduced  this  subject  to  me  
as well as Mr L. C. Kwek for many fruitful 
discussions.

\section*{References}

  1. F. Black and M. Scholes. " The Pricing of Options and Corporate Liabilities." 

Journal of Political Economy 81 ( May 1973), 637-59.

  2.  J.  C.  Hull. " Options, Futures and Other Derivative Securities", Prentice-

Hall (1993).

  3.  R.  C.  Merton.  "The  Theory  of  Rational Option Pricing." Bell Journal of 

Economics and Management Science 4 (Spring 1973), 141-83.

  M.  Garman.  "  A  General  Theory  of  Asset  Valuation  under  Diffusion State 

Processes." Working Paper No 50, University of California, Berkeley, 1976.

  4.  J.-P.  Bouchaud,  D. Sornette. " The Black-Scholes option pricing problem in 

mathematical  finance." Journal de Physique I 4, 863-81 (1994); Journal de Physique I 

5, 219-20 (1995). Journal de Physique I 6, 167-175 (1996).

J.-P. Bouchaud, G. Iori, D. Sornette, RISK 9, No 3, March 1996.

  5.  J.  C. Hull and A. White. " The Pricing of Options on Assets with Stochastic 

Volatilities." The Journal of Finance, Vol XLII, No. 2 (June 1987), 281-299.

J.  C. Hull and A. White. " An Analysis of the Bias in Option Pricing caused 

by  Stochastic  Volatility." Advances in Futures and Option Research, Vol3 (1988) 29-

61.

 E.  Stein  and  J.  Stein,  "  Stock  Price  Distribution  with  Stochastic  

Volatility: An Analytic Approach."  The Review of Financial Studies, (1991) 4, 727.

C.  Ball  and  A. Roma, " Stochastic Volatility Option Pricing."  Journal of 

Financial and Quantitative Analysis (1994) 29, 589.

S.  Heston,  " A Closed Form Solution for Options with Stochastic Volatility 

with  Applications  to  Bond  and Currency Options."  The Review of Financial Studies 

(1993) 6, 327.  Note this paper only considers the case of zero correlation (r = 0).

  6.  T.  Finucane  "  Binomial  Approximation of American Call Option Prices with 

Stochastic  Volatilities."  4th  Symposium  on  the  Frontiers  of Massively Parallel 

Computation, 1992, McLean, Virginia.

  7.  K.  Mills, M. Vinson, G. Cheng. " A Large Scale Comparison of Option Pricing 

Models  with  Historical  Market  Data."  NPAC  Technical  Report  SCCS-260, Syracuse 

University (1993)

  8.  D.  Duffie.  "  Dynamic  Asset  Pricing  Theory." Princeton University Press 

(1994).

  9.  R.  P.  Feynman  and  A.  R. Hibbs. " Quantum Mechanics and Path Integrals." 

McGraw-Hill (1965).

A. Das.  " Field Theory: A Path Integral Approach." World Scientific (1993).

  10.  J.  Dash.  "Path  Integral  and  Options -- I." Financial Strategies Group, 

Merrill Lynch Capital Market, 1989.

  11. M. S. Makivic. " Numerical Pricing of Derivative Claims: Path Integral Monte 

Carlo Approach." NPAC Technical Report SCCS 650, Syracuse University, (1994).

  12.  D.  W.  Heerman.  "  Computer  Simulation  Methods in Theoretical Physics." 

Springer Verlag (1990).

  13.  B.  E.  Baaquie  and L. C. Kwek  "Implied Volatility Smiles and Frowns from 

Stochastic Volatility."  Submitted for publication (1997).

\section*{Appendix 1  Black-Scholes Formula for Constant Volatility}

For  the  case of constant volatility the Black-Scholes equation is given by 
the following [2] 
$$
\frac{\partial f}{\partial t}= - \frac{1}{2} \sigma^2 S^2 
\frac{\partial^2 f}{\partial S^2} - rS \frac{\partial f}{\partial S} 
+ rf \eqno{(A1)}  
$$

Making the change of variable as in (2) 
$$
S = e^x,  \hspace{1cm} -\infty \leq x \leq \infty
$$

yields 
$$
\frac{\partial f}{\partial t} = (H + r)f
$$

with the Black-Scholes Hamiltonia given by
$$
H_{BS} = -\frac{\sigma^2}{2} \frac{\partial^2}{\partial x^2} + (\frac{1}{2} 
\sigma^2 - r) \frac{\partial}{\partial x} \eqno{(A2)}
$$

To  evaluate  the  price of the European call option with constant volatility 
we have the Feynman-Kac formula
$$
f(t,x) = \expon^{-r(T-t)}\int_{-\infty}^{\infty} dx^\prime < x |
\expon^{-(T-t)H_{BS}}|x^{\prime} > g(x^{\prime})
$$

To  compute  $p_{BS}(x, \tau|x^{\prime}) = < x | \expon^{-\tau H_{BS}} | x^{\prime} >$,
 where  remaining time is $\tau = T - t$,  
we  use  the Hamiltonian approach to illustrate another powerful technique of quantum
mechanics.   We  go  to  the  'momentum' basis in which $H_{BS}$ is diagonal.  The Fourier transform of the $|x> $ basis to 'momentum space' is given by
\bsq
<x | x^{\prime} >    =   \delta(x - x^{\prime})  &= & 
\int_{-\infty}^{\infty} \frac{dp}{2 \pi} \expon^{i p (x - x^{\prime})}\\
&=&  \int_{-\infty}^{\infty} \frac{dp}{2 \pi}  < x | p > < p | x^{\prime} >
\esq    
which yields for momentum space basis $| p> $ the completeness equation
$$
I =  \int_{-\infty}^{\infty} \frac{dp}{2 \pi} | p >< p | \eqno{(A3)}
$$    
and the scalar product
$$
< x | p >   =   \expon^{ipx};  \hspace{1cm} < p | x > =   \expon^{-ipx}.
$$

From (A2) and the equation above we have the matrix elements of $H$ given by
$$
<x|H_{BS}|p> = H_{BS}\expon^{ipx} = \{\frac{1}{2}\sigma^2 p^2 + i(\frac{1}{2}\sigma^2 - r)p\}\expon^{ipx}
$$

Using (A3) and the equation above yields
\bsq
< x|\expon^{-\tau H_{BS}}|x^{\prime} >  &=&  
\int_{-\infty}^{\infty} \frac{dp}{2\pi} 
< x | \expon^{-\tau H_{BS}} | p><p|x^{\prime}>  \\
&=& \int_{-\infty}^{\infty}\frac{dp}{2\pi} 
\exp(-\frac{1}{2} \tau \sigma^2 p^2) \expon^{ip(x-x^{\prime} + 
\tau (r -  \sigma^2 / 2))}    \\
&=& \frac{1}{\sqrt{2\pi\tau\sigma^2}} 
\exp [ -\frac{1}{2\tau\sigma^2} \{ x - x^{\prime} + \tau (r - \sigma^2 /
2) \}^2]\mbox{\hspace{3cm}}(A4)
\esq

 The  result above is the Black-Scholes distribution.  
Recall $x^{\prime} = \log (S(T)), x = \log(S(t))$ 
and $\tau = T - t$; 
the stock price evolves randomly from its given value of 
$S(t)$ at  time $t$ to a whole range of values for $S(T)$ at time $T$.  Equation (A4)
above states that $\log(S(T))$ has a normal distribution with mean equal to
$\log(S(t)) + (r-\sigma^2 /2)(T-t)$ and  variance  of
$\sigma^2 (T-t)$  as  is  expected for the Black-Scholes case with
constant volatility [2].

In  general for a more complicated (nonlinear) 
Hamiltonian such as the one given in  eqn.  (12b) for stochastic 
volatility it is not possible to exactly diagonalize $H$ and  then  
be able to exactly evaluate the matrix elements of $\expon^{-\tau H}$.
The Feynman path integral  is an efficient theoretical tool  
for analysing such nonlinear Hamiltonians, and this is the reason for using the path integral formalism.

\section*{Appendix 2  The Lagrangian}
    
The  lagrangian  $L$  is  central  to  the  path  integral  formulation of quantum mechanics.   We  first  find  $L_{BS}$ for  the Black-Scholes case of constant volatility before  tackling  the more complicated case.  For infinitesimal time $\epsilon$ it
 is given by Feynman's formula
$$
< x | \expon^{-\epsilon H_{BS}} | x^{\prime} >  =  
N(\epsilon) \expon^{\epsilon L_{BS}}
$$

where $N(\epsilon)$  is  a  normalization constant.
Since the formula (A4) is exact, we have, for $\delta x = x - x^{\prime}$
$$
L_{BS} = -\frac{1}{2\sigma^2}\{\frac{\delta x}{\epsilon}
+ r - \frac{1}{2} \sigma^2 \} ^2  \eqno{(A5a)}
$$

and with 
$$
N(\epsilon) = \frac{1}{\sqrt{2\pi\epsilon\sigma^2}} \eqno{(A5b)}
$$

We  now  analyze  the  case  of  stochastic  voltility.
From  eqn. (12b), the Hamiltonian is given by    
$$
H = -\frac{\expon^Y}{2} \frac{\partial^2}{\partial x^2}
+ (\frac{1}{2} \expon^Y - r) \frac{\partial}{\partial x} -
\xi \rho \expon^{Y/2} \frac{\partial^2}{\partial x \partial y}
- \frac{\xi^2}{2} \frac{\partial^2}{\partial y^2} +
(\frac{1}{2} \xi^2 - \mu) \frac{\partial}{\partial y} \eqno{(A6a)}
$$

Hence we obtain using (A3)
\bsq
<x, y| \expon^{-\epsilon H}| x^{\prime}, y^{\prime}> &=& \int_{-\infty}^{\infty} \frac{dp_X}{2\pi} \frac{dp_Y}{2\pi} < x, y | \expon^{-\epsilon H} | p_{X}, p_{Y}><p_{X}, p_{Y} | x^{\prime}, y^{\prime}> \\
&=& \int_{-\infty}^{\infty} \frac{dp_X}{2\pi} \frac{dp_Y}{2\pi}
\expon^{ip_X (x - x^{\prime})} \expon^{ip_Y (y - y^{\prime})}
\expon^{- \epsilon H (x, y, p_X, p_Y)}\mbox{\hspace{3cm}} (A6b)
\esq
and from (A4) the matrix elements of the Hamiltonian is given by
$$
H(x, y, p_X, p_Y) = \frac{\expon^{Y}}{2} p_x^2 + 
(\frac{1}{2} \expon^{Y} - r) ip_X + \xi \rho \expon^{Y/2} 
p_X p_Y + \frac{\xi^2}{2} p_Y^2  +  (\frac{1}{2} \xi^2 - \mu) ip_Y  
\eqno{(A7)}
$$

To  perform  the  Gaussian  integration  over  $p_X$ and $p_Y$  we need the inverse and the determinant of the matrix 
$$
M =  \left[
\begin{array}{ll}
\expon^{Y} & \xi \rho \expon^{Y/2} \\
\xi \rho \expon^{Y/2} & \xi^2 
\end{array}
\right]
\eqno{(A8)} 
$$
Note 
$$
{\rm det}M = \xi^2 \expon^{Y} (1 - \rho^2)    \eqno{(A9)}
$$
and 
$$
M^{-1} = \frac{1}{\xi^2 (1 - \rho^2)} 
\left[
\begin{array}{ll}
\xi^2 \expon^{-Y} & -\xi \rho \expon^{-Y/2} \\
-\xi \rho \expon^{-Y/2} & 1
\end{array}
\right]
\eqno{(A9)}
$$
Let
$$
A = x - x^{\prime} + \epsilon r - \frac{\epsilon}{2} 
\expon^{Y}   \eqno{(A11)}
$$
and
$$
B = y - y^{\prime} + \epsilon \mu - 
\frac{\epsilon}{2} \xi^2   \eqno{(A12)}
$$
From  (A6)  and  (A7),  on  performing the Gaussian integrations, we have the Feynman relation
$$
<x, y | \expon^{-\epsilon H} | x^{\prime}, y^{\prime}> = 
\frac{1}{2\pi \epsilon \sqrt{{ \rm det} M}} \expon^{\epsilon L}   \eqno{(A13)}
$$
where from (A6), (A8) and (A10)
$$
L = -\frac{1}{2 \epsilon^2 (1 - \rho^2)}(\expon^{-Y}A^2 +\frac{1}{\xi^2} B^2
-2 \frac{\rho}{\xi} \expon^{- Y/2} AB) + O(\epsilon) \eqno{(A14)}
$$
and  finally  from (A11) and (A12), for $\delta x = x-x^{\prime}$ and $\delta y = y-y^{\prime}$, we have the negative definite lagrangian given by
\bsq
L = & - & 
\frac{1}{2 \xi^2} ( \frac{\delta y}{\epsilon} + 
\mu - \frac{1}{2} \xi^2) ^2 \\
&-& \frac{\expon^{-Y}}{2(1-\rho^2)}
[\frac{\delta x}{\epsilon} + r  - \frac{1}{2} 
\expon^{Y} - \frac{\rho}{\xi} \expon^{Y/2} 
(\frac{\delta y}{\epsilon} + \mu - \frac{1}{2} \xi^2 )]^2 + 
O(\epsilon)  \mbox{\hspace{3cm}} (A15)
\esq

\section*{Appendix 3: The  $\int DX$ - Path Integration}    

The path integration over the $x(t)$-variables in eqn. (19) can be done exactly, and the derivation is given below [9].

Let
$$ Q = \int DX \expon^{S_X} \eqno{(A16)} $$   
where from (22b) the $x$-dependent term of the Lagrangian is given by
$$L_X(i) = -\frac{\expon^{-Y_i}}{2(1-\rho^2)}\{\frac{\delta x_i}{\epsilon} + r - \frac{1}{2} \expon{Y_i} - \frac{\rho}{\xi} \expon^{Y_i / 2}(\frac{\delta y_i}{\epsilon}
+ \mu - \frac{1}{2} \xi^2)\}^2 + O(\epsilon)    \eqno{(A16b)}$$
Let
$$ c_i = r - \frac{1}{2}\expon^{Y_i} -  \frac{\rho}{\xi} \expon^{Y_i / 2}
(\frac{\delta y_i}{\epsilon} + \mu - \frac{1}{2} \xi^2 )  \eqno{(A17a)} $$ 
Then
$$ S_X = - \frac{1}{2\epsilon(1-\rho^2)} \sum_{i = 1}^{N} \expon^{-Y_i}(x_i - x_{i - 1} + \epsilon c_i)^2   \eqno{(A17b)} $$
with boundary values given by
$$ x_0 = x^{\prime}, x_N = x  \eqno{(A17c)}$$
We make the change of variables
$$ x_i = z_i - \epsilon \sum_{j = 1}^{i} c_j,
\hspace{1cm} dx_i = dz_i, \hspace{0.5cm} i=1,2,\ldots,N-1
\eqno{(A18a)}$$
with boundary values
$$ z_0 = x^{\prime} \eqno{(A18b)}$$
$$ z_N = x + \epsilon \sum_{j = 1}^{N} c_j   \eqno{(A18c)}$$
We hence have from (A17b) and (A18a)
$$ S_Z = - \frac{1}{2\epsilon(1-\rho^2)} \sum_{i = 1}^{N} \expon^{-Y_i} (z_i - z_{i-1})^2   \eqno{(A19a)}$$
and, from (25c) and (A18a)
$$ 
Q = \frac{ \expon^{-Y_N /2}}{\sqrt{2\pi\epsilon(1-\rho^2)}}
\prod_{i = 1}^{N-1} \int_{-\infty}^{+\infty} \frac{d z_i \expon^{-Y_i /2}}{\sqrt{2\pi\epsilon(1-\rho^2)}}\expon^{S_Z}   \eqno{(A19b)}$$
All the $z_i$  integrations can be performed exactly; one starts from the boundary by first integrating over $z_1$ , and then over $z_2$, ....and finally over $z_{N-1}$.  Consider the $z_1$ -integration.  We have
$$
\int_{-\infty}^{+\infty} 
\frac{d z_1 \expon^{-Y_i /2}}{\sqrt{2\pi\epsilon(1-\rho^2)}} \exp \{
- \frac{1}{2\epsilon(1-\rho^2)}
[\expon^{-Y_2} (z_2 - z_1)^2 + \expon^{-Y_i}(z_1 - z_0)^2] \} $$
$$ \hspace{2cm}
= \frac{\expon^{Y_2 /2}}
{\sqrt{\expon^{Y_1} + \expon^{Y_2}}}
\exp \{ -\frac{1}{2\epsilon(1-\rho^2)} 
\frac{1}{\expon^{Y_i} + \expon^{Y_2}} (z_2 - z_0)^2 \}  
\eqno{(A20)} $$
Repeating this procedure $(N-1)$ times yields
$$ Q = \frac{\expon^{S_1}}{ \sqrt{2 \pi \epsilon(1-\rho^2) 
\sum_{i = 1}^{N} \expon^{Y_i}} }    \eqno{(A21a)} $$
where
\bsq
S_1 &=& - \frac{1}{2\epsilon(1-\rho^2) \sum_{i =1}^{N} \expon^{Y_i}} 
(z_N - z_0)^2 \\
&=& - \frac{1}{2\epsilon(1-\rho^2) 
\sum_{i =1}^{N} \expon^{Y_i}} \{ x - x^{\prime} +  
\epsilon \sum_{i = 1}^{N} c_i \}^2
\mbox{\hspace{7cm}} (A21b)
\esq

For the case of constant volatility $\xi = 0 = \rho$ and $\expon^{Y_i} = \sigma^2$ = constant for 
all $i$; eqn. (A21b) then reduces to the Black- Scholes given in eqn. (A3).


\section*{ Appendix 4 : Generating Function $Z (j,y,p)$}

We evaluate
$$ 
Z(j,y,p) = \int  DY \exp[  \int_0^\tau dt j(t)y(t)] \expon^{pY(0)/2}
\expon^{S_0} ;  \eqno{(A22a)}
$$
from (27b) we have
$$ S_0  = -\frac{1}{2 \xi^2} \int_0^\tau  
dt (\frac{dy}{dx} + \mu -  \frac{1}{2} \xi^2 )^2  \eqno{(Ab)}
$$
with boundary condition
$$ y(\tau) = y   .\eqno{(A22c)}
$$

We first evaluate $Z(j, y, y^\prime)$ 
given by the path integral in (A22a) but with the 
boundary condition
$$
y(0) = y^\prime   , y(\tau) = y    \eqno{ (A23)}
$$
and from eqn. (A22a)
$$
Z(j, y, p) = \int_{-\infty}^{\infty}   dy^\prime 
Z(j, y, y^\prime) \expon^{py^\prime/2}. \eqno{(A24)}
$$

Define new path integration variables $z(t)$ by
$$
z(t) = y(t) - y^\prime +  \frac{t}{\tau} (y^\prime - y). \eqno{(A25)}
$$

Note the new variables $z(t)$ due to (A23) have boundary conditions
$$
z(0) = 0 = z(\tau).    \eqno{(A26)}
$$

Hence
$$
Z(j, y, y^\prime) = \expon^{W_0} \int DZ \expon^{S_Z}   \eqno{(A27)}
$$
with
$$ 
W_0  = - \frac{1}{2\tau \xi^2}  (y^\prime - y - \mu \tau 
+  \frac{1}{2}\xi^2 \tau^2)^2  + 
y^\prime \int_0^\tau  dt j(t)
$$
$$
\mbox{\hspace{3cm}} - \frac{y^\prime - y}{\tau} \int_0^\tau  dt t j(t)
\eqno{(A28)}
$$
and 
$$
S_Z  = \int_0^\tau  dt j(t)z(t) -  \frac{1}{2 \xi^2} \int_0^\tau dt
(\frac{dz}{dt})^2.  \eqno{ (A29)}
$$

To perform the path integral over $z(t)$, 
note that from boundary condition (A26) we 
have the Fourer sine expansion
$$
z(t) = \sum_{n = 1}^{\infty} \sin(\pi n t/\tau) z_n  \eqno{(A30)}
$$

From (A29) and (A30) we have
$$
S_Z  = -\frac{\pi^2}{4 \tau \xi^2} \sum_{n = 1}^{\infty} 
n^2  z_n^2   +   \sum_{n=1}^{\infty} 
[ \int_0^\tau dt j(t) \sin(\pi n t/\tau) ] z_n
\eqno{(A31)}
$$

The path integration over $z(t)$ 
factorizes into infinitely many Gaussian integrations 
over the $z_n$ 's and we obtain
$$
\int  DZ \expon^{S_Z}   = C^\prime  \int_{-\infty}^{\infty} 
dz_1 dz_2 dz_3 \cdots dz_{\infty}  \expon^{S_Z} \eqno{(A32)}
$$
$$ \mbox{\hspace{2cm}}  = C(\tau) \expon^W \eqno{(A33)} $$
where
$$   W =  \frac{\xi^2 \tau}{\pi^2} \int_0^\tau dt dt^\prime 
j(t) D(t,t^\prime) j(t^\prime)    \eqno{(A34)}
$$
$$
\mbox{\hspace{3cm}} = \frac{\xi^2}{\tau} \int_0^\tau  dt 
\int_0^t dt^\prime j(t) (\tau - t) t^\prime j(t^\prime)    \eqno{(A35)}
$$
since
$$D(t,t^\prime) = \sum_{n = 1}^{\infty} \frac{1}{n^2}
\sin(\pi nt/\tau) \sin(\pi n t^\prime/\tau)
\eqno{(A36)}
$$
$$
\mbox{\hspace{3cm}}  = \frac{\pi^2}{2 \tau}   
[ t^\prime \theta(t - t^\prime) + t \theta (t^\prime - t) - \frac{t
t^\prime}{\tau}]   .   \eqno{(A37)}
$$
where the step function is defined by
$$
\theta(t) =  
\left\{ \begin{array}{ll}
1 & t > 0 \\
1/2 & t = 0 \\
0 & t < 0 
\end{array}
\right.    \eqno{(A37a)}
$$

The normalization function $C(\tau)$ 
can be evaluated by first considering the discrete 
and finite version of the $DZ$ - 
path integral along the lines discussed in Section 
III.  The result is given [Ref 9] by
$$ 
C(\tau) = \frac{1}{\sqrt{2\pi \xi^2 \tau}}\eqno{(A38)}
$$
Collecting our results we have 
$$    
Z(j, y^\prime, y) =  \frac{\expon^{W_0 + W}}{\sqrt{2 \pi \xi^2 \tau}} 
\eqno{(A39)}
$$
and performing the $y^\prime$ Gaussian integration in (A24) finally yields
$$
Z(j, y, p) = \expon^F \eqno{(A40)}
$$
with
$$
F = y \int_0^\tau  dt j(t) + (\mu -  \frac{1}{2} \xi^2 ) \int_0^\tau dt
(\tau - t)j(t)
$$
$$
\mbox{\hspace{5cm}} + \xi^2 \int_0^\tau  dt j(t) (\tau - t) \int_0^t
dt^\prime  j(t^\prime) + F^\prime   \eqno{(A41)}
$$ 
where
$$
F^\prime = \frac{1}{2} yp +  \frac{1}{2} \xi^2 p \int_0^\tau  dt 
(\tau - t)j(t) +  
\frac{1}{2} p\tau (\mu -  \frac{1}{2} \xi^2 ) + \frac{1}{8} p^2 \xi^2
\tau   .   \eqno{(A42)}
$$

\end{document}